\documentclass{jps-cp}
\usepackage{txfonts} %Please comment out this line unless the txfonts package is availabe in your LaTeX system.

\usepackage{color}
\def\bge{\begin{equation}}
\def\ene{\end{equation}}
\def\bg{\begin{eqnarray}}
\def\en{\end{eqnarray}}
\def\nn{\nonumber}

\def\d0bar{{\bar{D}^0}}

\def\Qbar{\overline{Q}}

\title{
In-medium Properties of the Low-lying Bottom
Baryons in the Quark-Meson Coupling Model
}

\author{Kazuo \textsc{TSUSHIMA}$^{1}$}

\inst{$^{1}$Laborat\'{o}rio de F\'{i}sica Te\'{o}rica e Computacional-LFTC, 
Universidade Cruzeiro do Sul, 01506-000, S\~ao Paulo, SP, Brazil}

\email{kazuo.tsushima@gmail.com}

\recdate{January 15, 2019}

\abst
{In-medium properties of the low-lying baryons 
are studied in the quark-meson coupling (QMC) model, 
focusing on the $\Sigma_b$ and $\Xi_b$ baryons.
It is predicted that the Lorentz-scalar effective mass of $\Sigma_b$ becomes 
smaller than that of $\Xi_b$ at moderate nuclear matter density, and as the density 
increases, namely, $m^*_{\Sigma_b} < m^*_{\Xi_b}$, although in vacuum $m_{\Sigma_b} > m_{\Xi_b}$. 
We also study the effects of the repulsive Lorentz-vector potentials on  
the excitation energies of these bottom baryons.
}

\kword{bottom baryons, in-medium properties, quark-meson coupling model}

\begin{document}
\maketitle

\section{Introduction}
Studies of in-medium baryon properties, especially for   
the baryons which contain charm and/or bottom quarks,      
are very interesting, due to the emergence of heavy-quark symmetry.  
In particular, in-medium properties of heavy baryons which contain  
at least one light $u$ or $d$ quark, can provide us with important 
information on the dynamical chiral symmetry, 
and the roles of light quarks in partial restoration 
of chiral symmetry. To study the in-medium properties of heavy baryons,   
we use the quark-meson coupling (QMC) model~\cite{Guichon:1987jp}, 
a quark-based model of nuclear matter, finite nuclei, and hadron properties 
in a nuclear medium~\cite{Saito:2005rv,Krein:2017usp}.
This report is based on the recent article~\cite{Tsushima:2018goq}. 
(See Refs.~\cite{Guichon:1987jp,Saito:2005rv,Krein:2017usp}
for details on the QMC model, and its successful applications.)

\section{Finite (hyper)nucleus in the QMC model}

Before discussing the in-medium baryon properties,  
we start with the case of finite (hyper)nucleus.
Using the Born-Oppenheimer approximation, 
a relativistic Lagrangian density which gives the same mean-field equations
of motion for a nucleus or a hypernucleus at the hadron level, may be given by    
the QMC model~\cite{Tsushima:2018goq,Tsushima:1997cu}:  
%%%
\begin{eqnarray}
{\cal L}_{QMC} &=& {\cal L}^N_{QMC} + {\cal L}^Y_{QMC},
\label{eq:LagQMC} \\
{\cal L}^N_{QMC} &\equiv&  \overline{\psi}_N(\vec{r})
\left[ i \gamma \cdot \partial
- m_N^*(\sigma(\vec{r})) - (\, g_\omega \omega(\vec{r})
+ g_\rho \dfrac{\tau^N_3}{2} b(\vec{r})
+ \dfrac{e}{2} (1+\tau^N_3) A(\vec{r}) \,) \gamma_0
\right] \psi_N(\vec{r}) \quad \nn \\
  & &\hspace{-10ex} - \dfrac{1}{2}[ (\nabla \sigma(\vec{r}))^2 +
m_{\sigma}^2 \sigma(\vec{r})^2 ]
+ \dfrac{1}{2}[ (\nabla \omega(\vec{r}))^2 + m_{\omega}^2
\omega(\vec{r})^2 ] %\nn \\
%
% & & 
+ \dfrac{1}{2}[ (\nabla b(\vec{r}))^2 + m_{\rho}^2 b(\vec{r})^2 ]
+ \dfrac{1}{2} (\nabla A(\vec{r}))^2, 
\label{eq:LagN} \\
{\cal L}^Y_{QMC} &\equiv&
\overline{\psi}_Y(\vec{r})
\left[ i \gamma \cdot \partial
- m_Y^*(\sigma(\vec{r}))
- (\, g^Y_\omega \omega(\vec{r})
+ g^Y_\rho I^Y_3 b(\vec{r})
+ e Q_Y A(\vec{r}) \,) \gamma_0
\right] \psi_Y(\vec{r}), 
\nn\\
& &\hspace{25ex} (Y = \Lambda,\Sigma^{0,\pm},\Xi^{0,-},
\Lambda^+_c,\Sigma_c^{0,+,++},\Xi_c^{0,+},\Lambda_b,\Sigma_b^{0,\pm},\Xi_b^{0,-}).
\label{eq:LagY}
\end{eqnarray}
%%%
In the above $\psi_N(\vec{r})$ and $\psi_Y(\vec{r})$
are, respectively, the nucleon (N) and hyperon (Y) fields. 
The mean-meson fields represented by, $\sigma, \omega$ and $b$ which 
directly couple to the light quarks self-consistently, are  
the Lorentz-scalar-isoscalar, Lorentz-vector-isoscalar and third component of  
Lorentz-vector-isovector fields, respectively, while $A$ is the Coulomb field.

In an approximation that the $\sigma$-, $\omega$- and $\rho$-mean fields couple
only to the $u$ and $d$ light quarks, the coupling constants for the hyperon 
appearing in Eq.~(\ref{eq:LagY})  
are obtained/identified as $g^Y_\omega = (n_q/3) g_\omega$, and
$g^Y_\rho \equiv g_\rho = g_\rho^q$, with $n_q$ being the total number of
valence light quarks in the hyperon $Y$, where $g_\omega$ and $g_\rho$ are 
the $\omega$-$N$ and $\rho$-$N$ coupling constants. $I^Y_3$ and $Q_Y$
are the third component of the hyperon isospin operator and its electric
charge in units of the proton charge, $e$, respectively.
The approximation used in the QMC model, that the meson fields couple only to 
the light quarks, reflects the fact that the magnitudes of the light-quark condensates 
decrease faster than those of the strange and heavier quarks 
as the nuclear density increases. 
This is associated with partial restoration of chiral symmetry 
in a nuclear medium (dynamically symmetry breaking and its partial restoration). 

The $\sigma$-field dependent 
$\sigma$-$N$ [$g_\sigma(\sigma(\vec{r}))$] 
and 
$\sigma$-$Y$ [$g^Y_\sigma(\sigma(\vec{r}))$] 
coupling strengths, associated with $m_N^*(\sigma(\vec{r}))$ in Eq.~(\ref{eq:LagN})  
and $m_Y^*(\sigma(\vec{r}))$ in Eq.~(\ref{eq:LagY}), respectively, are defined by 
\bg
& &m_N^*(\sigma(\vec{r})) \equiv m_N - g_\sigma(\sigma(\vec{r}))\, \sigma(\vec{r}),  
\label{effnmass}
\\
& &m_Y^*(\sigma(\vec{r})) \equiv m_Y - g^Y_\sigma(\sigma(\vec{r}))\, \sigma(\vec{r}),  
\hspace{3ex} (Y = \Lambda,\Sigma,\Xi, 
\Lambda_c,\Sigma_c,\Xi_c,\Lambda_b,\Sigma_b,\Xi_b), 
\label{effymass}
\en
where $m_N$ ($m_Y$) is the free nucleon (hyperon) mass. 
The $\sigma$-dependence of these coupling strengths must be calculated 
self-consistently within the quark model~\cite{Guichon:1987jp,Saito:2005rv}.

\section{Baryon properties in symmetric nuclear matter}

We consider the rest frame of   
symmetric nuclear matter, a spin and isospin saturated system 
with only strong interaction (Coulomb force is dropped) 
based on  Eq.~(\ref{eq:LagQMC}).
Within the Hartree mean-field approximation, 
the nuclear (baryon) $\rho_B$, and scalar $\rho_s$ densities 
are, respectively, given by,
\begin{equation}
\rho_B = \dfrac{4}{(2\pi)^3}\int d^3{k}\ \theta (k_F - |\vec{k}|)
= \dfrac{2 k_F^3}{3\pi^2},
%\label{rhoB}
%\\
%
\hspace{3ex} 
\rho_s = \dfrac{4}{(2\pi)^3}\int d^3 {k} \ \theta (k_F - |\vec{k}|)
\dfrac{m_N^*(\sigma)}{\sqrt{m_N^{* 2}(\sigma)+\vec{k}^2}} \, .
\label{rhos}
\end{equation}
Here, $m^*_N(\sigma)$ is the value (constant) of the Lorentz-scalar effective nucleon mass at 
a given nuclear density, and $k_F$ the Fermi momentum.

The Dirac equations in the standard QMC model (modeled by the MIT bag) for the quarks and antiquarks 
in nuclear matter, in a bag of a hadron, $h$ ($q = u$ or $d$ and $Q = s,c$ or $b$ hereafter) 
neglecting the Coulomb force, are given by  
[$x=(t,\vec{x})$ and for $|\vec{x}|\le$ 
bag radius]~\cite{Saito:2005rv}, 
\begin{eqnarray}
\left[ i \gamma \cdot \partial_x -
(m_q - V^q_\sigma)
\mp \gamma^0
\left( V^q_\omega +
\dfrac{1}{2} V^q_\rho
\right) \right] 
\left( \begin{array}{c} \psi_u(x)  \\
\psi_{\bar{u}}(x) \\ \end{array} \right) &=& 0,
\label{Diracu}\\
\left[ i \gamma \cdot \partial_x -
(m_q - V^q_\sigma)
\mp \gamma^0
\left( V^q_\omega -
\dfrac{1}{2} V^q_\rho
\right) \right]
\left( \begin{array}{c} \psi_d(x)  \\
\psi_{\bar{d}}(x) \\ \end{array} \right) &=& 0,
\label{Diracd}\\
\left[ i \gamma \cdot \partial_x - m_{Q} \right] \psi_{Q} (x) = 0, ~~~~~~~~  
\left[ i \gamma \cdot \partial_x - m_{Q} \right] \psi_{\overline{Q}} (x) &=& 0,  
\label{DiracQ}
\end{eqnarray}
where, the (constant) mean fields for a bag are defined by 
$V^q_\sigma \equiv g^q_\sigma \sigma$, 
$V^q_\omega \equiv g^q_\omega \omega$, and
$V^q_\rho \equiv g^q_\rho b$,
with $g^q_\sigma$, $g^q_\omega$ and
$g^q_\rho$ being the corresponding quark-meson coupling constants. 
The mass of the hadron $h$ in a nuclear medium, $m^*_h$, 
is calculated by~\cite{Guichon:1987jp,Saito:2005rv}
\begin{eqnarray}
m_h^* &=& \sum_{j=q,\bar{q},Q,\Qbar} 
\dfrac{ n_j\Omega_j^* - z_h}{R_h^*}
+ \frac{4}{3}\pi R_h^{* 3} B_p,\quad
\left. \dfrac{\partial m_h^*}
{\partial R_h}\right|_{R_h = R_h^*} = 0,
\label{hmass}
\end{eqnarray}
%%%%%%
where $\Omega_q^*=\Omega_{\bar{q}}^*
=[x_q^2 + (R_h^* m_q^*)^2]^{1/2}$, with
$m_q^*=m_q{-}g^q_\sigma \sigma=m_q-V^q_\sigma$, and 
$\Omega_Q^*=\Omega_{\Qbar}^*=[x_Q^2 + (R_h^* m_Q)^2]^{1/2}$,
and $x_{q,Q}$ are the lowest mode bag eigenvalues.

In Table~\ref{bagparambc} we present the model inputs in vacuum, 
the quantities calculated in vacuum and at normal 
nuclear matter density, $\rho_0 = 0.15$ fm$^{-3}$.

%%%%%%%%%%%%%%%%%%%%%%%%%%%%%%%%%%%%%%%%%%%%%%%%%%%%%%%
\begin{table}[tbh]
\begin{center}
\caption{
The parameters related with the zero-point energy $z_B$; 
baryon masses and the bag radii in free space
[at normal nuclear matter density, $\rho_0=0.15$ fm$^{-3}$]
$m_B$(MeV), $R_B$(fm) [$m^*_B, R^*_B$]; and the lowest mode bag eigenvalues $x_1, x_2, x_3$ 
[$x^*_1, x^*_2, x^*_3$] of baryon $B(q_1,q_2,q_3)$ with 
the corresponding valence quarks $q_1, q_2, q_3$ in the baryon $B$, 
where $z_B$'s are kept the same as those in vacuum, i.e., density independent.
Free space mass values $m_B$ for the heavy baryons from Ref.~\cite{PDG},   
those for the strange hyperons from Ref.~\cite{Saito:2005rv}, 
and the nucleon bag radius $R_N = 0.8$ fm (and $m_q=5$ MeV), are inputs.
The light quarks are indicated by $q = u$ or $d$.
Note that the baryons containing at least one light quark $q$  
are modified in the medium in the QMC model, but $\Omega, \Omega_c$, and $\Omega_b$ 
are not modified in the QMC model.
We recall that some inputs are updated from those in Refs.~\cite{Saito:2005rv,Krein:2017usp} 
based on the data~\cite{PDG}.
For the recent data for $\Sigma_b$, see Ref.~\cite{Aaij:2018tnn}, which gives the 
averaged mass of $m_{\Sigma_b} = 5813.1$ MeV, to be consistent with the value extracted 
from Ref.~\cite{PDG}. The entry with ``NA'' stands for ``not applicable''.
}
\label{bagparambc}
\vspace{-3ex}
\begin{tabular}{c|cccccc|ccccc}
\hline
\hline
$B(q_1,q_2,q_3)$ &$z_B$ &$m_B$ &$R_B$ &$x_1$ &$x_2$ &$x_3$ &$m_B^*$ &$R_B^*$ 
&$x^*_1$ &$x^*_2$ &$x^*_3$\\
\hline
$N(qqq)$           &3.295 &939.0  &0.800 &2.052 &2.052 &2.052   & 754.5 &0.786 &1.724 &1.724 &1.724\\
\hline
$\Lambda(uds)$     &3.131 &1115.7 &0.806 &2.053 &2.053 &2.402   & 992.7 &0.803 &1.716 &1.716 &2.401\\
$\Sigma(qqs)$      &2.810 &1193.1 &0.827 &2.053 &2.053 &2.409   &1070.4 &0.824 &1.705 &1.705 &2.408\\
$\Xi(qss)$         &2.860 &1318.1 &0.820 &2.053 &2.406 &2.406   &1256.7 &0.818 &1.708 &2.406 &2.406\\
$\Omega(sss)$      &1.930 &1672.5 &0.869 &2.422 &2.422 &2.422   &NA     &NA    &NA    &NA    &NA   \\
\hline
$\Lambda_c(udc)$   &1.642 &2286.5 &0.854 &2.053 &2.053 &2.879   &2164.2 &0.851 &1.691 &1.691 &2.878\\
$\Sigma_c(qqc)$    &0.903 &2453.5 &0.892 &2.054 &2.054 &2.889   &2331.8 &0.889 &1.671 &1.671 &2.888\\
$\Xi_c(qsc)$       &1.445 &2469.4 &0.860 &2.053 &2.419 &2.880   &2408.3 &0.859 &1.687 &2.418 &2.880\\
$\Omega_c(ssc)$    &1.057 &2695.2 &0.876 &2.424 &2.424 &2.884   &NA     &NA    &NA    &NA    &NA   \\
\hline
$\Lambda_b(udb)$  &-0.622 &5619.6 &0.930 &2.054 &2.054 &3.063   &5498.5 &0.927 &1.651 &1.651 &3.063\\
$\Sigma_b(qqb)$   &-1.554 &5813.4 &0.968 &2.054 &2.054 &3.066   &5692.8 &0.966 &1.630 &1.630 &3.066\\
$\Xi_b(qsb)$      &-0.785 &5793.2 &0.933 &2.054 &2.441 &3.063   &5732.7 &0.931 &1.649 &2.440 &3.063\\
$\Omega_b(ssb)$   &-1.327 &6046.1 &0.951 &2.446 &2.446 &3.065   &NA     &NA    &NA    &NA    &NA   \\
\hline
\hline   
\end{tabular}
\end{center}
\end{table}
%%%%%%%%%%%%%%%%%%%%%%%%%%%%%%%%%%%%%%%%%%%%%%%%%

\begin{figure}
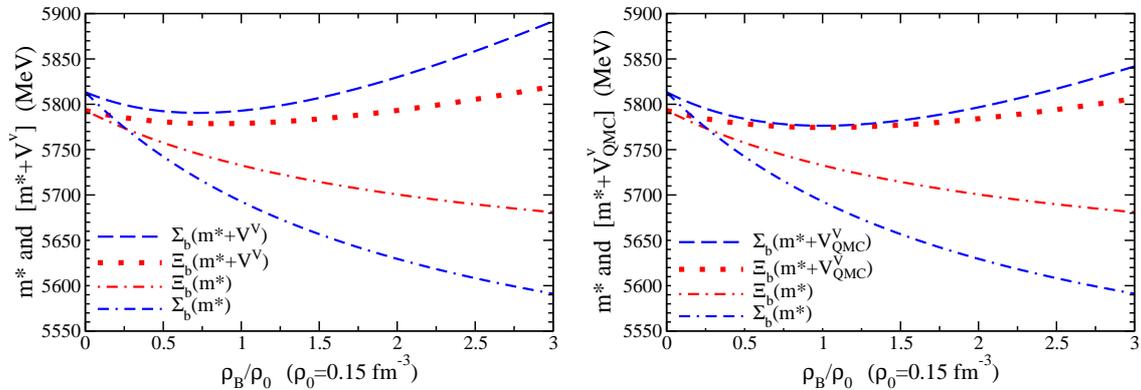

\centering
%\vspace{2.5ex}
\includegraphics[scale=0.3]{mattbc_SbXbEnergy.eps}
\hspace{1ex}
\includegraphics[scale=0.3]{mattbc_SbXbVqmc.eps}
\caption{\label{fig:SbXbEnergy} 
Lorentz-scalar effective masses ($m^*$), and excitation energies   
for two cases of vector potentials, 
with (left panel, denoted by $m^*+V^V$) and 
without (right panel, denoted by $m^*+V^V_{QMC}$) the phenomenologically introduced 
Pauli (vector) potentials~\cite{Tsushima:1997cu} based on Pauli Principle at the quark level.
}
\vspace{-2ex}
\end{figure}

We focus on the in-medium properties of $\Sigma_b$ and $\Xi_b$ baryons.
In Fig.~\ref{fig:SbXbEnergy} we show the Lorentz-scalar effective masses, and    
excitation energies (Lorentz-scalar effective masses plus 
vector potentials) for two cases of vector potentials.
%%%%%%%%%%%%%%
%%%%%%%%%%%%

First, one can notice in Fig.~\ref{fig:SbXbEnergy} that 
the Lorentz-scalar $\Sigma_b$ effective mass becomes smaller 
than that of $\Xi_b$, namely,  $m^*_{\Sigma_b} < m^*_{\Xi_b}$,  
at baryon density range larger than about $0.3 \rho_0$, 
although in vacuum, $m_{\Sigma_b} > m_{\Xi_b}$ (see Table~\ref{bagparambc}).
This is one of the highlighted predictions in the present study.

Next, we discuss the excitation energy, which is the sum of the Lorentz-scalar 
effective masses plus Lorentz-vector potential. 
The left (right) panel is the case with (without) the phenomenologically introduced 
Pauli (vector) potentials based on the Pauli Principle at the quark 
level~\cite{Tsushima:1997cu}. 
In the results shown in Fig.~\ref{fig:SbXbEnergy}, the realistic case should be 
without the Pauli potentials (right panel) for the present case of bottom baryons.  
Interestingly, for the realistic case without the Pauli potentials, 
the excitation energies for $\Sigma_b$ and $\Xi_b$ are nearly degenerate 
in the nuclear matter density range 
$0.5 \rho_0 < \rho_B < 1.5 \rho_0$.  
Namely, $\Sigma_b$ and $\Xi_b$ can be produced at rest with nearly 
the same cost of energy. This may imply the emergence of many interesting phenomena,  
for example, in heavy ion collisions and reactions in the systems of dense nuclear medium, 
such as in the deep core of a neutron (compact) star.
 
The results shown in Fig.~\ref{fig:SbXbEnergy} also suggest that  
the two different types of the vector potentials can possibly be distinguished, 
and give important information on the dynamical symmetry breaking 
and partial restoration of chiral symmetry.
For proving these suggestions, one needs to seek what kind of experiments can be made 
to get a clue. It might be very interesting to measure the valence quark (parton) distributions 
of $\Sigma_b$ and $\Xi_b$ in medium, since the supports of the parton distributions  
of these baryons reflect their excitation energies. Another possibility 
may be to measure the strangeness-changing semileptonic weak decay 
of $\Xi_b \to \Sigma_b$ in a medium, which reflects their excitation 
energy difference in a medium.

\section{Summary and conclusion}

We predict that the Lorentz-scalar effective mass of $\Sigma_b$ becomes 
smaller than that of $\Xi_b$ in the nuclear matter density    
range larger than $\simeq 0.3 \rho_0$ ($\rho_0 = 0.15$ fm$^{-3}$), 
while in vacuum the mass of $\Sigma_b$ is larger than that of $\Xi_b$.

We have also studied the effects of two different repulsive Lorentz-vector 
potentials on the excitation energies of $\Sigma_b$ and $\Xi_b$ baryons.
In the case without the phenomenologically introduced Pauli potentials, 
which is expected to be more realistic, the excitation energies 
for $\Sigma_b$ and $\Xi_b$ are predicted to be nearly degenerate 
in the nuclear matter density range about [$0.3 \rho_0, 1.5 \rho_0$]. 
Thus, the production of $\Sigma_b$ and $\Xi_b$ cost nearly the 
same energy at rest in this density range. 

To make possible connections of the findings for the Lorentz-scalar effective masses and/or 
excitation energies of $\Sigma_b$ and $\Xi_b$ baryons with experimental observables, 
one needs to seek relevant experimental methods and situations.
It might be very interesting to measure the valence quark (parton) distributions 
of $\Sigma_b$ and $\Xi_b$ in medium, since the supports of the parton distributions  
of these baryons reflect their excitation energies. Another possibility 
may be to measure the strangeness-changing semileptonic weak decay   
of $\Xi_b \to \Sigma_b$ in a medium, which reflects their excitation 
energy difference in a medium.

To conclude, studies of the $\Sigma_b$ and $\Xi_b$ properties in a nuclear medium, 
can provide us with very interesting and important information 
on the dynamical symmetry and partial restoration of chiral symmetry, 
as well as the roles of the light quarks in a medium.

%%%%%%%%%%%%%%%%%%%%%
%%%\subsubsection{Equation numbers}

%%%The \verb|seceq| option resets the equation numbers at the start of each section.
%%%%%%%%%%%%%%%%%%%%%

%%%%%%%%%%%%%%%%%
%%%\appendix
%%%\section{}

%%%Use the \verb|\appendix| command if you need an appendix(es). The \verb|\section| command should follow %%%even though there is no title for the appendix (see above in the source of this file).
%%%%%%%%%%%%%%%%

%%%%%%%%%%%%%%%%%%%%%%%%%%%%

%%%%%%%%%%%%%%%%%%%%%%%%%%%

\end{document}